\documentclass[sigconf,balance=false]{acmart}

\renewcommand\footnotetextcopyrightpermission[1]{}

\acmConference[]{}{}{}

\usepackage{defs}
\usepackage{pkgs}

\title{\textbf{Remote attestation of SEV-SNP confidential VMs using e-vTPMs}}

\date{}

\author{Vikram Narayanan}
\affiliation{\institution{University of Utah}
\city{Salt Lake City}
\state{Utah}
\country{USA}}

\author{Claudio Carvalho}
\affiliation{\institution{IBM Research}
\city{Yorktown Heights}
\state{New York}
\country{USA}}

\author{Angelo Ruocco}
\affiliation{\institution{IBM Research}
\city{Yorktown Heights}
\state{New York}
\country{USA}}

\author{Gheorghe Alm\'asi}
\affiliation{\institution{IBM Research}
\city{Yorktown Heights}
\state{New York}
\country{USA}}

\author{James Bottomley}
\affiliation{\institution{IBM Research}
\city{Yorktown Heights}
\state{New York}
\country{USA}}

\author{Mengmei Ye}
\affiliation{\institution{IBM Research}
\city{Yorktown Heights}
\state{New York}
\country{USA}}

\author{Tobin Feldman-Fitzthum}
\affiliation{\institution{IBM Research}
\city{Yorktown Heights}
\state{New York}
\country{USA}}

\author{Daniele Buono}
\affiliation{\institution{IBM Research}
\city{Yorktown Heights}
\state{New York}
\country{USA}}

\author{Hubertus Franke}
\affiliation{\institution{IBM Research}
\city{Yorktown Heights}
\state{New York}
\country{USA}}

\author{Anton Burtsev}
\affiliation{\institution{University of Utah}
\city{Salt Lake City}
\state{Utah}
\country{USA}}

\begin{document}

\maketitle

\thispagestyle{empty}

\section*{Abstract}

Trying to address the security challenges of a cloud-centric software deployment
paradigm, silicon and cloud vendors are introducing \textit{confidential
computing} -- an umbrella term aimed at providing hardware and software
mechanisms for protecting cloud workloads from the cloud provider and its
software stack.
Today, Intel Software Guard Extensions (SGX), AMD secure encrypted
virtualization (SEV), Intel trust domain extensions (TDX), etc., provide a way
to shield cloud applications from the cloud provider through encryption of the
application's memory below the hardware boundary of the CPU, hence requiring
trust only in the CPU vendor. 
Unfortunately, existing hardware mechanisms do not automatically enable the
guarantee that a protected system was not tampered with during configuration and
boot time.
Such a guarantee relies on a hardware root-of-trust, i.e., an integrity-protected
location that can store measurements in a trustworthy manner, extend them,
and authenticate the measurement logs to the user (remote attestation).

In this work, we design and implement a virtual trusted platform
module (\vtpm{}) that virtualizes the hardware root-of-trust 
without requiring trust in the cloud provider.
To ensure the security of a \vtpm{} in a provider-controlled environment,
we leverage unique isolation properties of the \snp{} hardware that allows
us to execute secure services (such as \vtpm{}) as part of the enclave
environment protected from the cloud provider. 
We further develop a novel approach to vTPM state management where the vTPM
state is not preserved across reboots.
Specifically, we develop a stateless \emph{ephemeral} \vtpm{} that supports
remote attestation without any persistent state on the host.
This allows us to pair each confidential VM with a private instance of a \vtpm{}
completely isolated from the provider-controlled environment and other VMs.
We built our prototype entirely on open-source components -- Qemu, Linux,
and Keylime.
Though our work is AMD-specific, a similar approach could be used to build
remote attestation protocols on other trusted execution environments~(TEE).

\section{Introduction}

Over the last two decades, public clouds have become an inescapable building block
of virtually every modern application. 
The move to the cloud created a unique security challenge. 
Both application vendors and end-users are required to trust the cloud 
infrastructure that is often in charge of handling security and privacy-sensitive data. 
Such trust is fragile as multi-tenant cloud environments are operated by third
party providers and include a large and complex virtualization and storage
stacks optimized for a wide variety of hardware and software execution
scenarios. 
Unfortunately, vulnerabilities in critical cloud software and infrastructure
are unavoidable.

In the last decade, three widely deployed virtual machine monitors (VMMs) --
Xen, KVM, and VMware -- that provide the foundation of isolation and security in the
cloud suffered from 428~\cite{cves:xen}, 111~\cite{cves:kvm}
and 154~\cite{cves:vmware} vulnerabilities each.
Cloud software stacks like Openstack and Cloudstack suffer from several
vulnerabilities, some resulting in total information disclosure and rendering
resources unusable~\cite{cves:openstack, cves:cloudstack}.
Moreover, physical access to the system opens the door for a range of hardware
attacks, e.g., memory extraction such as cold-boot~\cite{cold-boot-attack},
RAMBleed~\cite{rambleed, rambleed-openssl}, etc.

In an effort to minimize the TCB of cloud applications,
hardware vendors and some cloud providers have introduced support for
hardware-protected trusted execution environments
(TEEs)~\cite{intel-sgx-explained, wp:amd-sev, spec:intel-tdx, wp:arm-cca, ibm-pef}.
TEEs protect data in use from the host software stack including the hypervisor
and even the physical attacker. 
In effect, TEEs remove the cloud provider from the TCB, even though the
provider still manages the lifecycle of an application.

Isolation alone, however, is not sufficient to protect a workload or sensitive
data. 
To ensure integrity, modern systems rely on a combination of \emph{measured
boot}~\cite{mboot:uefi, mboot:bootstrapping-trust} and \emph{runtime
attestation}~\cite{ra:principles, ra:semantic}.
A measured boot protocol performs measurement of all binaries involved in the
boot of the system to ensure the integrity of all boot-time components, i.e., the platform
firmware, bootloader(s), and the operating system kernel.
Runtime attestation combines measured boot with integrity measurement architecture
(IMA) that ensures integrity measurements of all binaries loaded and executed by
the system after it booted, i.e., dynamic kernel extensions, system binaries, etc.
Attestation works by comparing entries in the measured boot and
IMA logs with a pre-defined set of acceptable values (called an
\emph{attestation policy}) and exposing any measurements that do not
conform to policy expectations.

Support for attestation requires a \emph{root-of-trust device}, i.e.,
an integrity-protected location that can store measurements in a
trustworthy manner, extend them, and authenticate the measurement logs to the user
(remote attestation).
On a physical machine, a trusted platform module (TPM) chip
can be used as the root of trust.
Some cloud providers offer virtual machines with virtual TPMs (vTPMS)
attached to them ~\cite{vtpm:gcp, vtpm:azure, vtpm:aws-nitro,vtpm:alibaba}.
These vTPMs, however, are emulated by the host
virtualization stack. Using this kind of emulated device requires
trusting the service provider, which is at odds with confidential computing.
In this paper, we show how to implement a confidential vTPM emulated inside
a TEE, isolated from both host and guest, linked to the root of trust of
the enclave, and providing similar properties to a physical TPM.

In this work, we design and implement a new virtual trusted platform module
(\vtpm{}) that virtualizes the hardware root-of-trust without requiring trust
in the cloud provider.  
To ensure the security of a \vtpm{} in a provider-controlled environment, we
leverage unique isolation properties of the \snp{} hardware that allows us to
execute secure services (such as \vtpm{}) as part of the enclave environment
protected from the cloud provider. 
We further develop a novel approach to the vTPM state management where the vTPM
state is not preserved across reboots.
Specifically, we develop a stateless \emph{ephemeral} \vtpm{} that supports
remote attestation without a persistent state on the host.
This allows us to pair each confidential VM with a private instance of a \vtpm{} that 
is completely isolated from the provider-controlled environment and other VMs.

We design our \vtpm{} around the following security requirements: 

\begin{itemize} 
	
\item 
\textbf{Isolation}: 
Physical TPMs are isolated at the hardware level.
Typical vTPMs emulated on the host are isolated from the guest via
		virtualization, but exposed to the trusted host.
In addition, the \vtpm{} also needs isolation from the guest operating system,
		since it acts as a root-of-trust device for attestation.  
A \vtpm{} should be isolated from both the host and the guest system.

\item 
\textbf{Secure communication}: In a physical TPM, communication is isolated at
		the hardware level, although these assurances can sometimes be
		subverted~\cite{tpm-genie, pr:linux-tpm2-hmac}.
In a typical \vtpm{}, the TPM commands and responses are transmitted through
		the untrusted hypervisor~\cite{vtpm:berger, vtpm:xen-doma,
		vtpm:xen-domb, vtpm:gvtpm, vtpm:xen-libos}.
An attacker can interpose on the channel and alter the request or response
		defeating the security guarantees offered by a
		TPM~\cite{tpm-genie}.
Communication with \vtpm{} should be secure. 

\item 
\textbf{Persistent state}: Physical TPMs have a persistent identity that is set
		when the device is manufactured. 
Maintaining persistent state in a virtualized environment usually requires a
		centralized management system to propagate and store \vtpm{}
		state. 
The management system is part of the TCB and is usually managed by the cloud
		provider. 
\vtpm{}'s state should be managed by the client and protected from 
		the cloud provider.

\end{itemize}

To implement isolation, we leverage unique properties of the \snp{} execution environment. 
Our confidential vTPM is emulated inside the \snp{} enclave (hence it is 
isolated from the host and the cloud provider). 
Moreover, we leverage Virtual Machine Privilege Levels (VMPLs) to isolate vTPM 
from the guest and hence ensure the integrity of remote attestation. 
Since our confidential vTPM is emulated inside the guest security context, the
guest and vTPM can communicate in plaintext without information being exposed
to the untrusted host.
Moreover, we ensure that neither the guest nor the hypervisor can tamper 
with the communication. 
To avoid exposing sensitive vTPM state to a complex management system, we
develop a new  ephemeral approach to vTPM state management, in which the 
state of the \vtpm{} never leaves the protected enclave.

The above security properties allow us to implement a vTPM that is
comparable in security and functionality to a physical TPM. Our vTPM
does not violate the trust model of confidential computing and extends
existing measurement capabilities to support sophisticated attestation
flows, enabling the creation of cloud-native workloads with a small
TCB that can be rigorously audited.

Our work leverages the unique architectural properties of the AMD \snp{} execution
environment; however, we will discuss how to generalize this solution at the end
of the paper. We will also expand on the properties of the ephemeral vTPM,
which does have certain restrictions. The limitations of an ephemeral vTPM
do not affect the attestation usecases described here.

Our contributions are as follows:
\begin{itemize}[leftmargin=*, nosep, labelindent=\parindent]
\item We propose using an ephemeral \vtpm{} to remove attacks to the \vtpm{} state.
\item We are the first to leverage the new features of AMD \sev{} to provide a secure implementation of a \vtpm{}.
\item We demonstrate a complete remote attestation workflow for our \svtpm{} solution, implicitly proving that remote attestation frameworks can provide
measured boot and remote attestation with an ephemeral \vtpm{}.

\end{itemize}

\section{Background and related work}

\subsection{Trusted execution environments}
Ubiquitous nature of cloud computing as a de facto large-scale application 
deployment paradigm resulted in a new security challenge -- protecting sensitive 
user data in a large, complex, and potentially untrusted environment of 
a cloud provider. 
To address the growing security concerns, a range of academic~\cite{sanctum} and
industry~\cite{intel-sdm,trustzone} projects proposed the idea of trusted execution
environments~(TEEs) in which the execution of a user program can be shielded from the 
rest of the software and hardware stack of the cloud provider.
TEEs provide isolated environments, or \textit{enclaves}, that ensure 
confidentiality and integrity of the user workload by relying only on the
processor.

Intel SkyLake architecture introduced software guard extensions
(SGX) that implement secure enclave for user-level applications through a 
combination of novel architectural extensions and CPU microcode.
SGX suffered from numerous vulnerabilities~\cite{sgx-attacks:survey},
ranging from access to the secrets inside the enclave to extracting the 
quoting enclave's attestation keys that allowed attackers to forge 
attestation reports~\cite{sgx-attack:foreshadow}.

In 2016, AMD introduced secure encrypted virtualization (SEV), where the
entire virtual machine --- as opposed to just part of an application --- 
was encrypted with an ephemeral key managed by a dedicated co-processor, 
AMD secure processor~(AMD-SP).
AMD-SP takes care of the lifecycle management of the \sev{}
VMs~\cite{spec:amd-sev} and serves as the integrated root-of-trust for the
AMD processor~\cite{wp:amd-sp}.
By using a unique key per VM, \sev{} isolates the guest VMs from the rest
of the host operating system and from other guests.

Intel trust domain extensions~(TDX) introduced their own
version of hardware-isolated encrypted virtual machines called trusted
domains~(TDs).
Intel TDX relies on an SGX-based quoting enclave called the TD-quoting
enclave to perform remote attestation of trusted domains~\cite{spec:intel-tdx}.
Unfortunately, the attestation keys used by the quoting enclave are long-lived, and when
leaked, affect millions of devices.

ARM introduced confidential compute architecture~(CCA) with their Armv9-A
architecture, where the processor provides an isolated hardware execution
environment called \emph{Realms}, for hosting entire VMs in a secure
space~\cite{wp:arm-cca}.
Similar to other TEEs~\cite{wp:amd-sev, spec:intel-tdx} ARM CCA provides
launch measurement for the realms and can do measured boot with their hardware
enforced security~(HES) module specification~\cite{spec:arm-cca-sec-model}
which serves as the root-of-trust~\cite{arm-cca:rss, arm-cca:rss-talk}.

\paragraph{AMD secure encrypted virtualization}

Since 2016, AMD has incrementally added additional protection features to SEV.
\seves{} (SEV encrypted state) protects the register state in the virtual
machine control block (VMCB) with encryption and integrity
protection~\cite{wp:sev-es}.
To communicate and share data with the hypervisor during hypercalls, guest
hypervisor communication block (GHCB) was introduced~\cite{spec:ghcb} that
would remain unencrypted.
Finally, with \snp{} (secure nested paging), AMD
introduced a reverse mapping table (RMP) which performs page validation and
keeps track of page ownership to prevent replay attacks~\cite{wp:sev-snp}.

\paragraph{Virtual machine privilege levels}
To avoid relying on the host infrastructure for running secure services for
the \cvm{}, AMD also introduced virtual machine privilege levels (VMPLs) in
\snp{}.
Similar to protection rings in x86 architecture, VMPLs allow a guest VM
address space to be subdivided into four levels with different privileges
(with \vmpl{0} being the highest privilege level).
Implementing privilege isolated abstraction layers with \vmpl{} allows the
design and deployment of secure services that are completely isolated from
the untrusted host operating system and the guest VM~\cite{wp:sev-snp}.

To standardize the communication between various services offered by the
software running at \vmpl{0} and the guest operating system AMD introduced
a specification called Secure VM service module
(\svsm{})~\cite{spec:sev-svsm}.
The protocol uses registers to pass the arguments and return values.
In the absence of \svsm{} firmware, the entire guest VM can execute under
\vmpl{0} unmodified.
However, with \svsm{}, they run at a lower privilege level,
corresponding to a higher VMPL~(i.e., 1-3), and require interaction with the
\svsm{} for some privileged operations.

\subsection{Integrity}

TEEs ensure confidentiality of the workload but do not 
guarantee integrity.
The trusted platform module~(TPM) is used along with a TEE to implement 
a secure root-of-trust in hardware. 
A TPM measures and records the cryptographic hash of the software during 
the boot process and reliably verifies the same at a later point in time.
TPM is implemented as a cryptographic co-processor chip that is embedded on
the motherboard of a platform.
It provides several cryptographic operations (e.g., encryption, signing,
hashing) and secure storage for small data such as keys.

\paragraph{Measured boot}

Measured boot is the process of recording the measurements of all boot
components during the system initialization process.
Hashes of all components are recorded in a log file that is authenticated
using the TPM.
This authentication works by extending TPM's Platform Configuration Registers (PCRs)
with digests of individual events in the boot log.
A TPM-signed \emph{quote} is used to vouch for the accuracy of the log.

\paragraph{Runtime integrity}

Integrity measurement architecture~(IMA) is a Linux subsystem that implements the idea 
of measured boot after the system is booted, e.g., measures hashes of all kernel extensions
before they are executed~\cite{ima}.
Together with measured boot, IMA enables a remote attestation protocol to ensure the
runtime integrity of the system.
Specifically, it allows an outside observer to ascertain specific
properties of a set of devices/machines.
As an example, one might be interested to ascertain the booted kernel, on a
set of machines in a data center. These properties of interest are
cumulatively called an attestation policy.
To ensure the integrity of the measurements, IMA relies on the TPM, i.e., extends the measurements 
into the TPM PCRs, similar to the measured boot log.

Measured boot and remote attestation are designed to stop an attacker who has control 
over the boot sequence of a system, e.g., an untrusted cloud provider, or an attacker who 
gains administrative privileges and can load malicious kernel extensions, 
or downgrade security critical subsystems to exploitable versions. 
These mechanisms complement a number of security mechanisms aimed to prevent
runtime exploitation of the system through a range of low-level
vulnerabilities~\cite{chen+:apsys11}, e.g., stack
canaries~\cite{cowan+:usenixss98}, address space randomization
(ASLR)~\cite{shacham+:acmccs04}, data execution prevention
(DEP)~\cite{exec-shield}, superuser-mode execution and access
prevention~\cite{smep, smap}, and even control-flow~\cite{intel-sdm} and
code-pointer integrity~\cite{cpi}.

\paragraph{Virtual trusted platform module (vTPM)}

A \vtpm{} is a pure software implementation of a TPM module as defined by
the TPM 2.0 specification~\cite{spec:tpm2.0}.
\vtpm{} enables the virtualization of a hardware root of trust across multiple
entities, i.e., virtual machines, and is aimed at providing functionality
identical to a hardware TPM.
Berger et al.~\cite{vtpm:berger} proposed the first design for virtualizing a
TPM that can be used for providing TPM functionalities to virtual machines.
Their design consists of a \vtpm{} manager and a set of \vtpm{} instances,
where the \vtpm{} manager executes as part of the VMM and takes care of
multiplexing physical hardware across multiple VMs. 
Berger et al. extend the TPM command specification to include support for
creating virtual instances and rely on hardware TPM for establishing trust.

Stumpf et al.~\cite{vtpm:hw-virt} proposed a virtual TPM design by applying
hardware virtualization techniques from Intel VT-x technology.
Their multi-context TPM contains different modes of execution and has a
dedicated TPM control structure for every VM, which would be loaded
by the VMM before invoking the TPM commands.
Several \vtpm{} architectures were proposed over the years: from a
generalized vTPM~\cite{vtpm:gvtpm} to separating \vtpm{} functionalities
across Xen domains with different privileges~\cite{vtpm:xen-libos,
vtpm:xen-doma, vtpm:xen-domb}.
Unfortunately, existing designs either place trust on the host environment (VMM, host
OS) or rely on the hardware TPM for establishing trust.
None of those designs satisfies the security and
confidentiality requirements of confidential computing.
Recent \vtpm{} designs move their implementation inside a TEE such as Intel
SGX~\cite{eTPM, vtpm-for-cloud, svtpm, cocotpm}.
Though this design offers protection from the cloud provider, the state of
the TPM must be securely stored and should be protected against rollback
attacks. Additionally, to avoid substitution attacks, both the vTPM and
the consuming VM must securely identify each other before services can be
provided.

\paragraph{Cloud \vtpm{}s}

Cloud providers that offer \cvm{}s typically provide virtual TPM device that
can serve as a root-of-trust and can also be used for remote
attestation.
Google cloud offers plain \sev{} \cvms{} and measured boot
attestation via a \vtpm{} managed by the
hypervisor~\cite{vtpm:gcp-shielded-vms}.
Microsoft Azure cloud relies on Azure attestation service for attesting
\cvms{}~\cite{vtpm:azure} that generates a token to decrypt the \vtpm{}
state and the disk.
Alibaba cloud offers \vtpm{} support on their elastic compute service
VMs~\cite{vtpm:alibaba}.
Amazon AWS provides Nitro TPM, a virtual TPM implementation conforming to
the TPM 2.0 specification as part of their EC2
offering~\cite{vtpm:aws-nitro}.
Some of these providers use a qemu-backed \vtpm{} that runs on the host,
and requires trust in the cloud provider.
Additionally, there is very limited public knowledge about the design and implementation 
of the above cloud \vtpm{}s what limits understanding of their security guarantees.
In contrast, our work results in an openly available \svtpm{} implementation
that is built on top of other standard opensource components~(i.e., Qemu,
Linux, and Keylime).
As our \svtpm{} relies only on the hardware-protected isolation environment
offered by the AMD-SP hardware, it allows cloud users to leverage our \vtpm{} as 
\svsm{} firmware and hence completely eliminate the need for trusting the cloud provider.

\begin{table*}[!]
  \small
  \caption{Feature comparison of \svtpm{} and other TEE-based \vtpm{}s}
  \vspace{-1em}
\begin{tabular}{cccccc}
  \hline
          & Trust anchor           & Persistent State      & Secure communication & Rollback protection & TPM management \\
  \hline\hline
  SvTPM~\cite{svtpm}    & SGX -\textgreater vTPM & Encrypted (SGX-Seal)   & SSL                  & Yes     & Self-contained            \\
CocoTPM~\cite{cocotpm}   & Self-signed sub-CA     & Encrypted (on disk)   & SSL                  & Yes    & Central             \\
SVSM-vTPM & AMD -\textgreater vTPM & Ephemeral             & Secure by design     & N/A    & Self-contained        \\
\hline
\end{tabular}
\label{table:relwork}
\end{table*}

\paragraph{TEE-based \vtpm{}s}

\autoref{table:relwork} presents a summary of differences between our
\svtpm{} design and other TEE-based \vtpm{}s.
\cocotpm{} proposes a unified architecture for attestation of
\cvms{} where the hypervisor launches a \cvm{} that acts as a \vtpm{}
manager and handles all the \vtpm{} instances~\cite{cocotpm}.
Several other projects rely on running \vtpm{} under isolation provided by
other hardware TEE mechanisms such as Intel SGX~\cite{svtpm, eTPM,
vtpm-for-cloud} and ARM Trustzone~\cite{fTPM}.
SvTPM aims to protect against NVRAM replacement, and rollback
attacks~\cite{svtpm} by running the \vtpm{} inside an SGX enclave for
KVM-based VMs, whereas
eTPM manages several enclave \vtpm{}s in a Xen environment and relies on a
physical TPM to provide root-of-trust~\cite{eTPM}, similar to Berger et
al.~\cite{vtpm:berger}.

To estasblish root of trust, SvTPM relies on Intel SGX datacenter
attestation primitives (DCAP) mechanism whereas \cocotpm{} uses a
self-signed certificate with which they sign the EK.
\svtpm{} establishes a chain of trust by generating an \snp{} attestation
report by passing the $digest(EK_{pub})$ as the user-data along with the
attestation request and thus relying only on the AMD hardware.

Both SvTPM and \cocotpm{} persists the state of the TPM.
SvTPM leverages SGX sealing to tie the persistent state of the TPM to the
appropriate VM whereas \cocotpm{} stores the state encrypted on the host
such that it can only be decrypted by the \cocotpm{}.
In contrast, by implementing an ephemeral \vtpm{}, we completely eliminate
the classes of attacks that come with state protection and endpoint
substitution.

Both \cocotpm{} and SvTPM require modifying parts of the software stack to
implement transport layer security (TLS) for securing the communication
channel between a VM and its \vtpm{}.
However, \svtpm{} implements an interface where both the command request
and response are part of the encrypted VM pages and hence secure by design.

To manage the state machine of the \vtpm{} instance and to maintain
the association betweeen a VM and its \vtpm{}, SvTPM and \cocotpm{} take
different approaches.
SvTPM follows a decentralized model where each \vtpm{} instance is hosted
on a separate SGX enclave whereas 
\cocotpm{} employs a central \vtpm{} manager where multiple \vtpm{}
instances are hosted on the same \cocotpm{} \cvm{}.
Though the \cocotpm{} VM is running inside a \cvm{}, a central design
suffers from several attacks ranging from denial of service to colluding
with other \cvm{}s.
Though it is possible to launch a dedicated \cocotpm{} for every
\cvm{}, it results in wastage of architectural resources as the
number of address space identifiers~(ASIDs) are limited.
In contrast, our \svtpm{} architecture equips each \cvm{} with their own
private \vtpm{} instance by leveraging the \svsm{} architecture that
implements VM privilege levels.
We propose a minimalistic \vtpm{} design that avoids the need to 
support secure communication and management of persistent state.
Also, by having a self-contained design and a simple API interface for
performing remote attestation, we avoid the complexities that are
associated with orchestrating a remote attestation
protocol~\cite{intel-dcap}.

\section{Threat model} 

We assume that an attacker has physical access to the machine and
unrestricted privileges on the software and firmware executing on the 
host machine, i.e., firmware, hypervisor and virtualization stack, and 
the host operating system. 
However, the memory of the confidential VM is protected by the AMD SEV 
technology, i.e., encrypted with a key only known to the AMD secure processor~(AMD-SP).
We trust the AMD hardware and the implementation of \snp{} and \svsm{}.

Ciphertext side channel attacks~\cite{sev-attack:cipherleaks,
sev-attack:cipherleaks-2} on the \sev{} encrypted VM (by building a
dictionary of plaintext-ciphertext pairs) are out of scope.
Attacks against the integrity measurement architecture~(IMA) such as
TOCTOU~\cite{ima:subverting-ima}, other measurement gaps such as code
injected by extended berkeley packet filter~(eBPF) are out of scope.
Also, runtime attacks exploiting stack or heap overflows such as
return-oriented programming on the guest VM are out of scope as IMA
measures only the persistent files.

\begin{figure}
  \centering
  \includegraphics[scale=0.65]{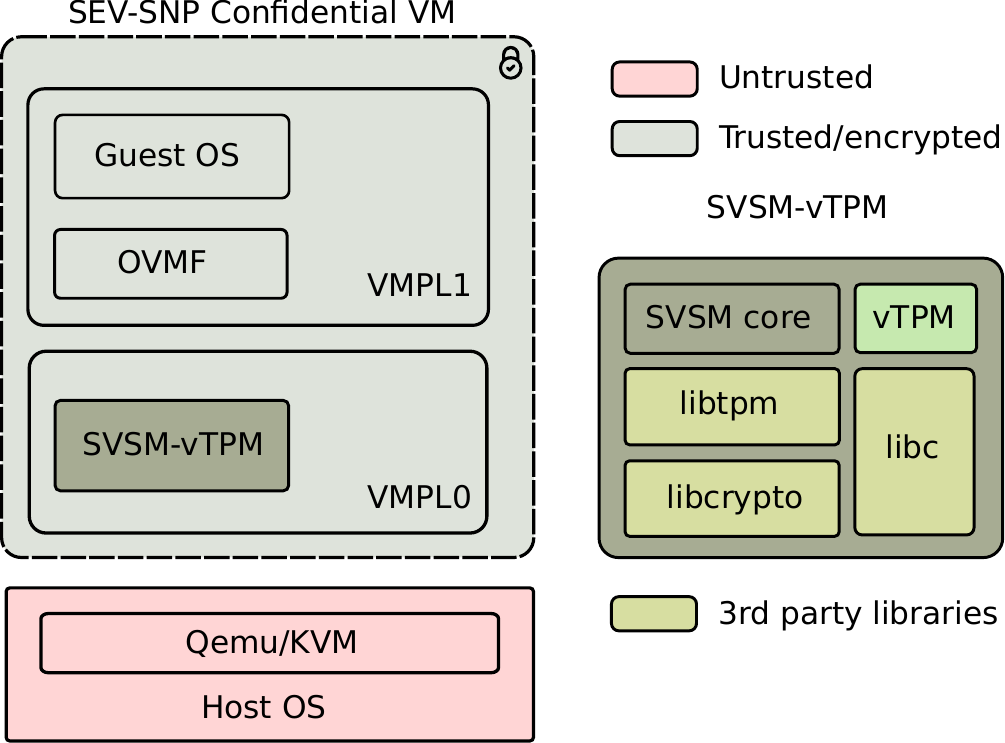}
  \caption{\svtpm{} architecture and its components}
  \label{fig:arch}
  \vspace{-4mm}
\end{figure}

\section{TPM virtualization with \svsm{}}

\svtpm{} is a secure virtual \tpm{} designed to enable remote attestation and
runtime integrity measurement in a provider-controlled confidential computing
environment backed by an AMD SEV hardware. 
Specifically, we do not trust any software on the host machine. 
To achieve strong isolation from the host, we leverage unique capabilities of
AMD SEV environment and execute a virtual TPM instance along with the guest
system inside a hardware-protected TEE enclave (\autoref{fig:arch}).
The entire \snp{} \cvm{} memory is encrypted by the AMD-SP.
\svtpm{} runs inside the VM privilege level 0 (\vmpl{0}), which allows us to both isolate
it from the rest of the guest system and provide secure communication between
the guest and the TPM. 
Specifically, we load a minimal baremetal execution environment in VMPL0 when a
new confidential VM is created.
Finally, we completely eliminate the burden of TPM state management such as
preserving the state, injecting it to the correct \cvm{} during boot-up,
and also prevent a whole class of attacks based on exfiltration of the TPM
state with a novel idea of an ephemeral TPM -- our TPM instances have no
persistent state to save or guard against.

\subsection{Isolation}

As \vtpm{} offers a virtual root-of-trust for the virtual machine, it has to be
hosted in an environment that provides strong isolation of its state and is 
designed to minimize the attack surface for a potential attacker.
Arguably, two design flaws undermine the security of existing \vtpm{}s to
be used in a confidential computing environment.

First, until recently, the cloud provider was a de facto part of the trust
domain.
\vtpm{}s were often managed and implemented as a
component inside the hypervisor~\cite{vtpm:berger} or as a part of the
virtualization stack~\cite{vtpm:xen-doma, vtpm:xen-domb, vtpm:xen-libos}.
To reduce the attack surface on the component hosting the \vtpm{}, several
alternative \vtpm{} architectures were proposed. 
Triglav \vtpm{} utilized dynamic root of trust~(DRTM) as a mechanism to ensure
the integrity of the hypervisor~\cite{triglav}. 
Another \vtpm{} solution utilized x86 system management mode~(SMM) for
isolation and protection of the TPM~\cite{vtpm:smm}.
Though such designs offer some form of protection against a non-malicious
cloud environment, they do not satisfy the requirements of confidential
computing where the entire host environment is untrusted.
Recent TEE-based \vtpm{}s run the \vtpm{} manager and several instances in
a hardware isolated TEE such as SGX~\cite{eTPM, svtpm, vtpm-for-cloud}, AMD
SEV \cvm{}~\cite{cocotpm} or in ARM Trustzone~\cite{fTPM}.

Second, historically, virtualization of TPM relied on a centralized
architecture.
The core part of the \vtpm{}, a \vtpm{} manager, responsible for instantiating
a TPM, multiplexing the communication between multiple VMs and \vtpm{}s, and
saving the TPM state in a secure location was shared across all \vtpm{}
instances~\cite{vtpm:berger, vtpm:xen-libos, vtpm:xen-doma, vtpm:xen-domb,
cocotpm}.
As the manager handles the lifecycle of all \vtpm{}s on a machine and has
access to the physical TPM hardware, it naturally becomes a central point for
attack.
A malicious VM can launch attacks ranging from a simple denial-of-service to
sophisticated attacks trying to exfiltrate the secrets by exploiting the
vulnerabilities in a centralized \vtpm{} manager.
If exploited, the security of all the \vtpm{}s handled by the manager is
compromised.

\begin{figure}
  \centering
  \includegraphics[scale=0.6]{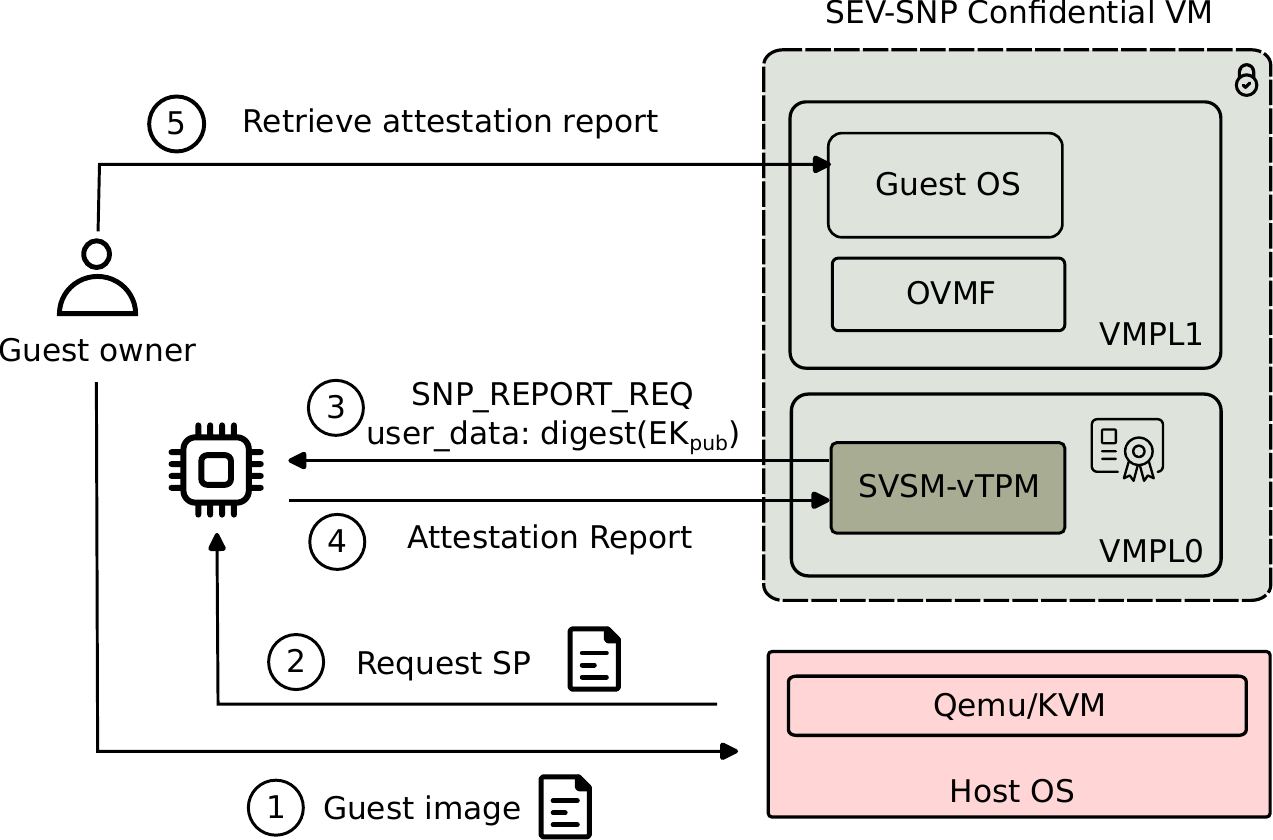}
  \caption{Generating \snp{} attestation report inside \svtpm{}}
  \label{fig:attestation}
  \vspace{-4mm}
\end{figure}

\paragraph{Private, isolated TPMs}

Instead of relying on a central \vtpm{} manager that manages several instances
of \vtpm{} in an untrusted environment, we base our design on two insights.
First, to provide strong isolation of the \vtpm{} code, we leverage the
architectural support offered by AMD \sev{}.
Second, to avoid centralized management, we rely on \svsm{} specification
that offers a way to implement secure services inside the guest VM.

Specifically, to ensure isolation, we leverage the VM privilege levels
inside the \cvm{} address space provided by the \svsm{} specification as
part of the \snp{} architecture.
In our architecture, every \cvm{} has its own private \vtpm{} that runs at
a higher privilege level (i.e., \vmpl{0}) inside each \cvm{} and is
encrypted by AMD-SP and has the same isolation guarantees of an encrypted
VM.

By running our \vtpm{} within an isolated privilege level within the guest
address space, we eliminate all the attacks that could be mounted on the
component that runs the \vtpm{}.
Additionally, operating at the \vmpl{0} offers additional protection that
it cannot be interfered by the guest or the host OS.

We use Qemu/KVM environment for running the \cvm{}.
\autoref{fig:attestation} shows how a \cvm{} is launched.
A user provides the boot-time binaries~(typically \svsm{} and OVMF) to be
loaded as part of the guest image~(\circlednum{1}).
Qemu communicates with the KVM which communicates with the \sev{} firmware
running inside the AMD-SP through an API interface to create a
\cvm{}~(\circlednum{2}).
The \svsm{} firmware is placed in \vmpl{0} and the OVMF firmware and the
rest of the guest environment~(i.e., the kernel and initrd in case of
direct boot) is placed at \vmpl{1}.

Unlike a regular programming environment that provides operating system
abstractions~(e.g., syscalls, timers, etc) and feature-rich libraries,
\svsm{} firmware runs on a restrictive bare-metal environment without
access to such features.
Enclave environments often come with such restrictions, for instance, one
would need a sophisticated library OS~\cite{sgx:graphene} to run unmodified
applications inside SGX.
However, in a bare-metal environment such as \svsm{}, one does not have
operating system abstractions such as timers/clocks, availability of crypto
libraries, etc.
However, a \vtpm{} needs to have access to timers, random numbers, and
cryptographic libraries for realizing a software TPM module.
We manually port the necessary libraries to satisfy the dependencies of the
TPM module.
Due to the encrypted code pages and the lack of interfaces between the
debugger and Qemu to install breakpoints inside the encrypted pages, 
we had to rely on print statements during development for debugging.

\subsection{Secure communication between VM and \vtpm{}} \label{subsec:secure-comm}
The communication channel between a VM and a corresponding \vtpm{} is a potential 
target for a range of security attacks, e.g., by altering the TPM command
requests and response buffers it is possible to subvert measured boot and runtime
attestation protocols~\cite{tpm-genie}. 
One way to mitigate such attacks is to secure the communication channel by 
implementing standards such as TPM HMAC and encryption~\cite{spec:tpm2.0} or 
DMTF secure protocol and data model~(SPDM) specification~\cite{spec:spdm}.
Though the TPM specification describes encryption and HMAC security layers,
there are very few TPM implementations that support them.
Developing a complex secure communication protocol such as
SPDM requires a large engineering effort. 
Recent \vtpm{} designs that rely on a hardware-protected TEEs implement
secure communication channel using transport layer security (TLS)
protocol~\cite{gh:openssl}.
Unfortunately, even a standard TLS protocol negatively affects the TCB size
of the TPM.

\paragraph{Secure communication}
Instead of implementing a secure communication protocol, we rely on the 
mechanism provided by AMD SEV and its ability to pass execution between virtual 
machine privilege levels.
While the transition between \vmpl{1} and \vmpl{0} triggers an exit into an
untrusted hypervisor controlled by the cloud provider, the internals of the
message remain protected inside the hardware-encrypted memory. 
Moreover, AMD SEV specification ensures that the hypervisor can only resume
execution of the VM at a corresponding privilege level, i.e., \vmpl{0}, if
the guest system triggers an exit into the hypervisor.
Hence, the hypervisor is unable to suppress messages unless the whole VM is
halted.

We rely on a generic platform device to interact with the \svtpm{} which
simply uses a page in memory for communicating the request and response
between the \cvm{} and the \svtpm{}~\cite{rfc:tpm-svsm-jejb}.
The guest kernel triggers an exit into the hypervisor, after
every write to the TPM command page.
Upon re-entry, the hypervisor puts the vCPU in \vmpl{0} where \svtpm{}
handler looks for TPM command ready flag and inturn invokes
the appropriate TPM command API to formulate the response buffer.
Then, the vCPU exits into \vmpl{1} and continues with the execution
of the guest VM.
We also make modifications to the TPM driver in OVMF to interact with our
\svtpm{}.

\subsection{\vtpm{} state}

A discrete physical TPM stores all the persistent state of the module inside
the chip's non-volatile (NV) store which holds the seeds for generation of
endorsement key (EK), storage root key (SRK) and also retains other values such
as NV Index values, objects made persisted by the TPM user, and state saved
when TPM is shutdown.
The TCG specification requires a TPM implementation to have some amount of
non-volatile storage for the operation of the TPM~\cite{spec:tpm2.0}.

As opposed to a physical TPM where the state of the TPM is securely stored
inside the TPM hardware chip inside a non-volatile RAM (NVRAM), a \vtpm{}
must manage its state in software.
Software \vtpm{}s typically implement the NV store in a disk-backed
file~\cite{vtpm:berger, vtpm:xen-libos, vtpm:xen-doma, vtpm:xen-domb, fTPM,
svtpm, cocotpm}.
Along with the software that implements the \vtpm{}, this NVRAM file is
part of the trusted computing base.
When a \vtpm{} is first initialized, the state file has to be created
on-the-fly or loaded from a file that is pre-created.

However, the state stored in the file needs to be secured against tampering and
rollback attacks~\cite{tpm:nvram-rollback}.
This could be achieved by encrypting the NV store file such that it could
be decrypted only by the \vtpm{} module.
This design calls for securely storing the secret key used to
encrypt/decrypt the NV state and inject it as a secret during the boot-up
of \vtpm{} module.
This brings in several complexities in the context of confidential
computing as the secret could only be injected during the launch phase.
First the user has to verify the \emph{launch measurement} of the load-time
components~(i.e., firmware, OVMF, etc.) before delivering the encrypted TPM
state along with the key to decrypt the TPM state.
The booting of the platform is blocked, waiting for the user to inject the
secret.
Additional care has to be taken to not give up the state to a \cvm{} that
is under the control of an attacker.

\paragraph{Ephemeral \vtpm{}}

Instead, our design choice of using an ephemeral \vtpm{} is much more
simplistic and pragmatic.
The \vtpm{} goes through the manufacturing process to generate a fresh set
of seeds, keys on every boot.
We avoid all the problems of handling persistent state, injecting it on
every boot, and guarding the encrypted state file by designing an ephemeral
\vtpm{} with no state.
First, ephemeral \vtpm{} is simple to implement: the NV storage becomes a
volatile storage and does not preserve any values across power cycles.
Second, it does not require any form of secrets to boot-up the \vtpm{} and the
platform.
Though there are downsides to this design such as: secrets cannot be
preserved across reboots, this offers much more flexibility as there is no
secret to guard against the aforementioned attacks.
Moreover, the programming environment for \svsm{} is extremely constrained in
terms of capabilities.
To save the TPM state on shutdown and to load the state on a reboot, the
\svsm{} should implement additional software to encrypt and decrypt the
state file.

\subsection{\svtpm{} provisioning}  \label{subsec:vtpm-provisioning}
After launching the \cvm{}, the hypervisor first loads and executes the
\svsm{} binary in \vmpl{0}.
Our modified \svsm{} follows the standard manufacturing process of
instantiating a \vtpm{} instance as specified by the TPM2.0
specification~\cite{spec:tpm2.0}.
First, we create a new endorsement key (EK) pair $\langle EK_{pub},
EK_{priv}\rangle$ from random seeds.
However, we do not create an endorsement key certificate ($EK_{cert}$) or a
platform certificate, as there is no entity to sign these certificates.

A significant, and much under discussed problem in Confidential
Computing is seeding the random number generator.
A VM when it boots has no natural sources of entropy that are not under the
control of the untrusted host.
In an ordinary VM, the x86 instructions \lstinline{RDRAND} and
\lstinline{RDSEED} cause \lstinline{VMEXIT}s; however, in \cvm{}s, these
instructions are guaranteed to provide direct access to the CPU hardware
random number in a way that the host cannot influence.
We use these instructions as the initial random number entropy source for
generating the random seeds.

\subsection{Adding vTPM to the trust chain}
Since our \svtpm{} module is instantiated with random seeds and does not
come with a manufacturer's certificate to verify the identity of the TPM,
we need to ensure the following security properties:
\begin{enumerate}[label=\textbf{S\arabic*}, nosep, labelindent=\parindent, leftmargin=*]
  \item Certify that the \svtpm{} is running in a real \cvm{} on genuine AMD hardware
  \item Certify that the \vtpm{} module is not tampered with.
  \item Communicate $EK_{pub}$ in a secure, tamper-proof way.
\end{enumerate}
To ensure these security properties, we rely on the attestation report from
the AMD-SP hardware.

\paragraph{\snp{} attestation report}
Software running at any \vmpl{} level can request an attestation report by
sending a message to the \sev{} firmware running inside the AMD-SP.
The request structure contains the \vmpl{} level and 512-bits of space for
user-provided data which would be included as part of the
attestation report signed by the AMD hardware.

\autoref{fig:attestation} shows the steps involved in getting an
attestation report.
On receiving a request to launch a VM, the platform loads the image and
cryptographically measures the contents of the image (\circlednum{1}).
Once the guest image is launched, the hypervisor
puts the vCPU in \vmpl{0} mode passing control to the \svsm{}
firmware~(after \circlednum{2}).
The \svsm{} firmware initializes the guest CPU, memory and sets up a
pagetable for execution and finally instantiates a \vtpm{}.
The \vtpm{} is provisioned as described
in~\autoref{subsec:vtpm-provisioning}.
Then, the \vtpm{} module requests an attestation report by sending a
\lstinline{SNP_REPORT_REQ} message to the AMD-SP hardware
(\circlednum{3}).
We place the digest of the public part of the generated endorsement
key~(i.e., $EK_{pub}$) in the user-data field of the request to communicate
the identity of the TPM to the guest VM.
The request message is encrypted with the appropriate VM platform
communication key (VMPCK) for that \vmpl{} level and prepended with a
message header which is integrity protected with authenticated
encryption~(AEAD).
The AMD-SP hardware decrypts the message, verifies the integrity and
responds with an attestation report(\circlednum{4}) that contains the
\emph{launch measurements}, vmpl level and the user-data (i.e.,
$digest\left(EK_{pub}\right)$).
We write this report into the NVIndex where the TPM would normally place
its EK certificate.
We can retrieve the saved attestation report at any point in time
(\circlednum{5}) as long as the guest VM is operational.
If needed, the guest VM can also place a report request to the AMD-SP
hardware from other \vmpl{} levels to generate a new attestation report.

\paragraph{Ensuring \textbf{S1}}
We can easily verify \textbf{S1} because the attestation report is
generated by the AMD-SP processor and signed using AMD's versioned chip
endorsement keys~(VCEK)~\cite{spec:amd-vcek}.
Verifying that the attestation report is genuine implicitly guarantee that
we obtained it from a genuine AMD processor, within a \cvm{}.

\paragraph*{Ensuring \textbf{S2}}
Before launching the \cvm{}, the AMD-SP hardware measures all the load-time
binaries as part of the \emph{launch measurement}. This includes the \svsm{}
and our \svtpm{} code.
By verifying these measurements that are included as part of the
attestation report, we can ensure that our \svtpm{} binary, and anything
else running in \vmpl{0}, has not been tampered.

\paragraph{Ensuring \textbf{S3}}
By verifying that the report request originated from \vmpl{0}, we can
confirm that the report was requested by a legitimate \svtpm{}, based on
\textbf{S2}.
By including the $digest(EK_{pub})$ as part of the attestation report~(via
user-data field), we offer a tamper-proof way to communicate the identify
of the TPM~($EK_{pub}$) to the entities interacting with this specific
\vtpm{}.
Since $EK_{pub}$ and $EK_{priv}$ are generated from random seeds provided
by the hardware~(i.e., \lstinline{RDRAND} and \lstinline{RDSEED}), as long
as the generator is tamper-proof, no entity can recreate $EK_{priv}$ and
impersonate this \vtpm{}.

\section{Implementation}

We base our implementation on the software stack recommended by AMD
which is publicly available on github~\cite{github:amdese}.
It consists of qemu, open virtual machine firmware (OVMF) and Linux
kernel for both the host and the guest all of which are modified to
support the AMD \snp{} architecture and will eventually be upstreamed.
We make minor modifications to the open-source framework Keylime~\cite{gh:keylime} for
performing remote attestation of VMs that has the SVSM-vTPM in the root of trust.

To implement \svtpm{}, we extend the open-source SVSM
implementation~\cite{github:amdese/linux-svsm} with a minimal C library~(a
stripped-down version of Musl~\cite{musl}), \libssl{}
library~\cite{gh:wolfssl} for cryptographic primitives, and Microsoft's TPM
that provides a software reference implementation of TCG's TPM 2.0
specification~\cite{gh:ms-tpm-ref}.

\subsection{Software TCB}
We add 1500 lines of code to the existing SVSM implementation in
Rust.
To implement \vtpm{}, we utilize third party libraries: a minimal C
library, WolfSSL crypto library and Microsoft's reference TPM
implementation~\cite{gh:ms-tpm-ref}.
The software TCB of our implementation is very similar to that of a
physical TPM consisting of a processor core (e.g., ARM SecureCore) that can
host the software components such as crypto libraries and TPM state
machine.
Also, the APIs we expose is similar to a hardware TPM implementing a CRB
interface.

We measure the \svsm{} code and other third-party crates that are part of
the dependency chain~(i.e., recursive dependencies) as everything is open
source.
We also assume that WolfSSL and Microsoft's TPM implementations are bug-free.
It should be noted that the microsoft reference implementation TPM is
also the code which is running in firmware inside a hardware TPM. For
this reason, we expect our vTPM to have the same security
characteristics as a hardware TPM: state exfiltration is prevented by
the VPML0 SNP security so the only attack vector is via the TPM
command interface.

\begin{figure*}
  \centering
  \includegraphics[scale=0.75]{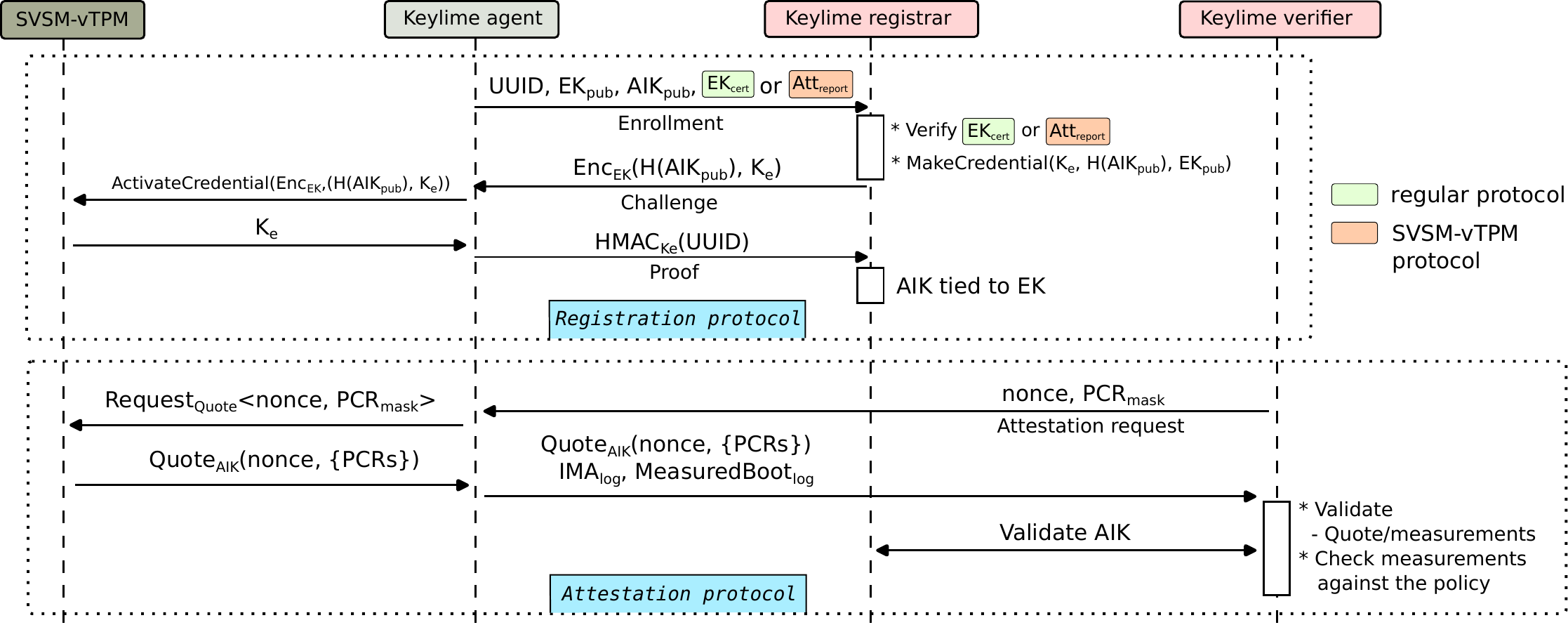}
  \caption{Remote attestation of a confidential VM using keylime and \svtpm{}}
  \label{fig:remote-attestation}
  \vspace{-4mm}
\end{figure*}

\subsection{Remote attestation with Keylime}

We use the Keylime package for remote attestation. Keylime is designed to
perform both boot-time and runtime attestation on a fleet of systems,
using the attested nodes' TPM devices as the root of trust~\cite{gh:keylime}.
The Keylime architecture is comprised of three major components: A
Keylime \emph{agent} is installed on every attested node. The agent
announces itself with a Keylime \emph{registrar} when it starts up.
The Keylime \emph{verifier} is in charge of performing attestations on
every node.
\paragraph{Registration protocol}
The purpose of Keylime registration is to record the availability of
the registering agent for attestation and to establish mutual trust
between the agent and the registrar. To this end the agent's
credentials are checked and an attestation key is negotiated between
the agent and the registrar for use for subsequent attestation
challenges.
As shown in ~\autoref{fig:remote-attestation}, the agent initiates the
enrollment process by sending its TPM credentials - i.e., the public
part of its endorsement key (EK) and attestation identity key (AIK),
as well as the EK certificate, and the node's UUID to the
registrar. The registrar verifies that the TPM's identity and
authenticity using the public EK and the EK certificate. Next, the
validity of the AIK is established through the
\emph{MakeCredential/ActivateCredential} function pair by using a
carefully constructed secret that can only survive the registrar to
agent roundtrip when both the TPM, AIK and UUID are authentic.
Identity verification of a normal TPM device involves checking that
the EK certificate correctly signs the public EK, and furthermore that
the EK certificate ($EK_{cert}$) is signed by a trusted root (such as
a manufacturer key or an intermediary key).

\paragraph{Attestation protocol}
Having successfully registered with the \emph{registrar}, the
\emph{agent} is now ready to service attestation challenges. The
Keylime \emph{verifier} initiates the attestation protocol by sending
a TPM quote request to the \emph{agent}, containing a nonce (to guard
against replay attacks) and a PCR mask (list of PCRs).
The \emph{agent} sends back the requested quote signed by the TPM,
using the AIK associated during the registration phase. In addition, a
number of logs (e.g. measured boot log, IMA log) are sent back with
the quote. The \emph{verifier} validates the TPM quote by decrypting
it with the registered AIK; validates the logs by testing them against
the PCRs contained in the quote; and finally checks the contents of
the logs against the attestation policy to render a
trustworthy/untrustworthy verdict.

\paragraph{Protocol changes to handle \svtpm{}s}

Since Keylime is built around interaction with TPM devices, we needed to make
only minor modifications in the code to handle \svtpm{}s. Basically, we only
had to modify how the Keylime verifier checks the authenticity of a TPM device
(function \lstinline{check_ek}).
As mentioned above, a ``normal'' TPM device is authenticated through
its EK certificate, which signs the public EK and in turn is verified
by a manufacturer certificate. Keylime carries a list of acceptable
manufacturer certificates, and any TPM in use by Keylime has to be
signed by one of these.
Our ephemeral \svtpm{}, by its very nature, is not provisioned with an
EK certificate. However, the (ephemeral) public EK is signed by the SEV
attestation report, which we validate by checking it against the
platform manufacturer's signature (i.e., AMD).
In order to minimize the required changes in Keylime, we decided to
simply replace the EK certificate with an SEV attestation report
($Att_{report}$) in our \svtpm{} (that is, we reuse the NVIndex in the
TPM where the EK certificate normally resides). The \emph{agent} reads
and submits the attestation report instead of the EK certificate
during registration. The modified \emph{registrar} validates the
attestation report (ensuring that it is signed by an authentic AMD
platform) instead of the validating the EK certificate~(marked by a
different color in ~\autoref{fig:remote-attestation}).
No other parts of the registration/attestation protocols require
changes for correct Keylime function.

\section{Evaluation}

We ran all our experiments on publicly-available cloudlab
infrastructure~\cite{cloudlab}.
We utilize a Dell Poweredge R6525 server equipped with two AMD EPYC 7543
32-core processor and 256 GiB RAM.
The host machine runs a 64-bit Ubuntu 20.04 Linux with v5.19 kernel and
qemu v6.1.50, whereas the confidential guest VM runs a 64-bit Ubuntu 22.04
Linux with a v5.17 kernel with open virtual machine firmware (OVMF) version
\lstinline{edk2-stable202208}, all of which are modified to enable
\snp{}~\cite{github:amdese}.
We have also evaluated our software stack on a Lenovo ThinkServer equipped
with an AMD EPYC 7763 64-core processor and 128 GiB RAM.

\subsection{Performance analysis} To understand the overheads of commonly
used TPM functionalities, we study the performance of several TPM commands on
\svtpm{} and compare that with a vanilla virtual machine that utilizes a
\vtpm{} hosted by Qemu.
We rely on Qemu/KVM to launch both the regular and \cvm{}.
Qemu-\vtpm{} setup uses the native TPM CRB interface as its frontend with
an \lstinline{swtpm} backend where the backed communicates with the \vtpm{}
running on the host userspace via a UNIX socket interface.
The \svtpm{} setup uses a generic platform driver~\cite{rfc:tpm-svsm-jejb}
to communicate with the \vtpm{} inside the \svsm{}~(as discussed in
\autoref{subsec:secure-comm}) running under \vmpl{0}.

We compare the performance of four different TPM commands
which are essential for remote attestation, i.e., \lstinline{PCRREAD},
\lstinline{PCREXTEND}, \lstinline{TPM2_QUOTE}, \lstinline{CREATEPRIMARY}.
These TPM commands briefly do the following:
\begin{itemize}[labelindent=\parindent, leftmargin=*, nosep]
\item{\textbf{PCR read}} This command reads the platform configuration
registers of the TPM.
A TPM may maintain multiple banks of PCR, where each bank is a collection
of PCRs extended with a specific hashing algorithm~(e.g., sha1, sha256).
In our benchmark, we read all the PCR values from all the banks~(i.e.,
sha1, sha256, sha384).

\item{\textbf{PCR extend}} performs an extend operation on a specific PCR from
a bank, i.e., it computes the hash of the old PCR value concatenated with
the input data, i.e.,
\newline
$PCR_{new} = hash(PCR_{old} || input\_data)$.
We extend a single PCR register from a sha256 bank.

\item {\textbf{Quote}} A TPM quote contains a subset of PCRs from a bank and a
nonce (to prevent replay attacks) signed by the attestation key~(AIK) of
the TPM.
We request a quote of three PCRs~(16-18) from two different banks~(sha1 and
sha256).

\item{\textbf{Create primary}} The TPM command creates a primary object under
the chosen hierarchy~(Endorsement, Platform, Owner or NULL) and loads it
into the TPM.
The TPM returns only a context with which one can interact with this object
and the public and private portions of the key are not returned.
We create an ECC keypair with the default curve~(ecc256).
\end{itemize}

\begin{figure}[t] \centering
  \includegraphics[width=1.0\columnwidth]{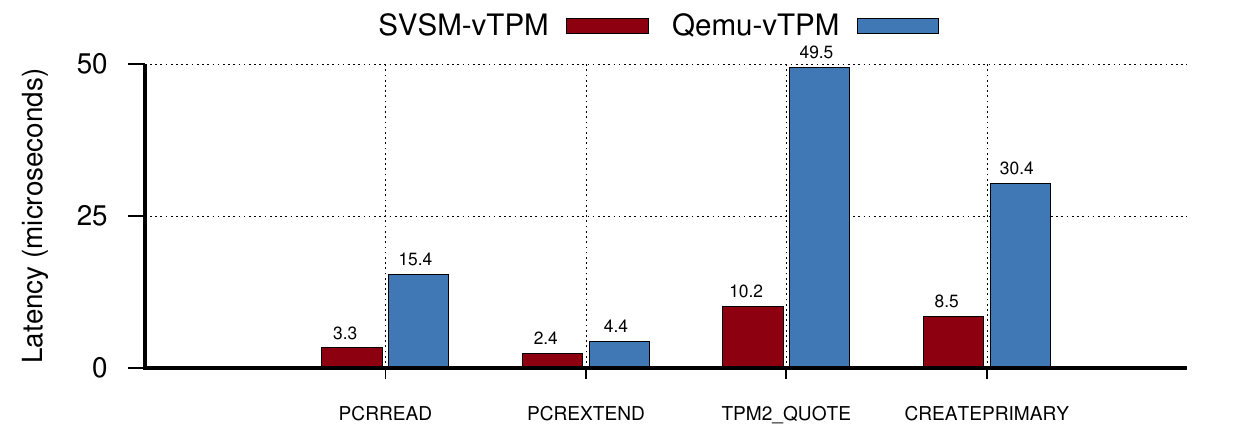}
  \caption{Performance overhead of \svtpm{} vs Qemu-\vtpm{}}
  \label{fig:tpm_overhead}
  \vspace{-4mm}
\end{figure}

We perform all the experiments by booting the \cvm{} with the corresponding
setup~(Qemu or \svsm{}), and invoke the TPM commands from the guest user
space using the \lstinline{tpm2-tools} package~\cite{gh:tpm2-tools}.
For each TPM command, we ran the benchmark for 3000 iterations.
We ran these experiments three times to measure the average
latency~(\autoref{fig:tpm_overhead}).
\svtpm{} incurs 5x lower latency than \qvtpm{} on \lstinline{PCRREAD}s and
to get a \lstinline{TPM2_QUOTE}.
We incur 1.8x and 3.5x lower latency on \lstinline{PCREXTEND} and
\lstinline{CREATEPRIMARY} TPM operations respectively.
Both qemu-hosted \vtpm{} and \svtpm{} incur an exit into the hypervisor to
communicate with the TPM.
However, our \svtpm{} suffers from much less overhead compared to the
qemu-hosted \vtpm{} as the latter involves communicating with the TPM
emulator backend (i.e., \lstinline{swtpm}) through the socket interface.

For completeness, we also ran the same experiments on a machine that has an
on-board physical TPM 2.0 device~(Nuvoton NPCT75x).
On an average, \lstinline{TPM2_QUOTE} and \lstinline{CREATEPRIMARY} is
25,000 times slower compared to our emulated \svtpm{} at 262,143 $\mu$s and
192,918 $\mu$s respectively whereas \lstinline{PCRREAD} is 9,000 times
slower (30,026 $\mu$s), and \lstinline{PCREXTEND} is 3,900 times slower
(9,359 $\mu$s).
In general, physical TPMs are an order of magnitude slower than emulated
ones because they are often connected to the mainboard via a low-bandwidth
bus such as serial peripheral interface~(SPI).

\subsection{Security Analysis}

A regular physical TPM is fully-isolated from the CPU and has its own
crypto engine, TPM state machine and a secure RNG inside the chip.
Moreover they do not store any of the TPM secrets on the DRAM and are not
vulnerable to memory attacks.
The gist of our security argument is that we are tying an ephemeral
\vtpm{} to the AMD-SP hardware's root of trust to perform runtime
attestation.
In this section we examine a number of potential security attacks that are
impossible to perform with a physical TPM and explain how our \vtpm{}
design prevents them.
Our hypothetical attacker's goal
would be to infiltrate and alter a guest \cvm{} without being detected
by the remote attestation system (Keylime).

\paragraph{Fake \vtpm{}}
The guest \cvm{} boots with the \svsm{} firmware containing our \svtpm{} as
part of the VM launch process.
The essence of this attack is that after the system is booted and the
keylime \emph{agent} is registered, an attacker could spawn a new
software \vtpm{} in the guest userspace to hijack all the \vtpm{}
commands and redirect to the newly spawned \vtpm{}.
The new fake software \vtpm{} is no longer running at a higher
privilege level and can be controlled by the attacker to forge TPM
quotes in an attempt to authenticate fake boot and IMA logs, and
therefore hide unauthorized software alterations from keylime.

However, once the registration protocol is complete, the keylime
registrar has associated the $EK$ of our ephemeral \vtpm{} with the
$AIK$ that would be used for signing the TPM quote.
With the above redirection of TPM commands to a fake vTPM an attacker
would not be able to forge the TPM quote, as the fake vTPM has no
access to the private $AIK$ of the original vTPM, safely hidden by
\vmpl{0} in \svsm{}.

The attacker could possibly force the registration protocol to restart
where an attacker could feed the TPM credentials from the newly
created \vtpm{}.
Again, keylime would detect this because of the mismatch of the fake TPM's
$EK_{pub}$ with its digest in the attestation report. A fake attestation
report cannot be generated because the report contains the \vmpl{} of
the entity that requested it, and the guest is not running at
\vmpl{0}.

\paragraph{Fake \snp{} attestation report}
We save the attestation report requested by \svtpm{} at the same NVIndex as
the $EK_{cert}$ to make it available to the keylime agent.
The essence of this attack is that the attacker could overwrite this NVIndex with
either garbage data or another attestation report after compromising the guest.
Garbage data would be detected by the keylime \emph{registrar},
resulting in attestation failure.
When overwritten with a genuine attestation report, an attacker can
potentially change the identity of the \vtpm{}, i.e., create another
\vtpm{}~(similar to Fake \vtpm{} attack) with a new set of keys and record
the new $EK_{pub}$ as part of the user-data field of the attestation
report.
If successful, they can perform all the attacks mentioned under the "Fake
\vtpm{}" attack~(i.e., spoof PCRs, forge quotes, etc).

Even though one could retrieve an attestation report from a different VM
privilege level, the platform guarantees that no one could spoof the
\vmpl{} level in the attestation report as it could be generated only by
the software running inside \vmpl{0}~(i.e., the keys for
encrypting the request message is available only at the corresponding
level).
Thus, the replaced attestation report, if valid, would contain a \vmpl{}
level greater than 0.
To prevent this attack, we check the \vmpl{} level while validating the
attestation report to ensure the requester \vmpl{} level is set to zero.

An attacker can overwrite the attestation report NVIndex with a genuine
attestation report off another \cvm{} or from a previous boot of this
\cvm{}.
Though the attestation report is signed by the AMD hardware, the user-data
will not match with the digest of $EK_{pub}$ we have inside the \svtpm{},
making the attack detectable.

\paragraph{Confidential VMs with no SVSM}
Though \vmpl{} levels are supported in the \snp{} specification, it cannot
be enforced by the end-user on a provider-controlled environment.
A malicious cloud provider could host a regular \sev{} VM and pretend
that it is running with an \snp{} firmware.
In this scenario, the \cvm{} would run without the \svsm{} firmware,
where the entire guest operating system will run under \vmpl{0}.
This makes it possible for a guest VM to generate its own attestation
report where the requester \vmpl{} level is set to 0.
To prevent this attack, the user could verify that the \cvm{} is booted with
the \svsm{} firmware running at \vmpl{0}
by measuring the boot-time binaries that includes the \svsm{} firmware
running at \vmpl{0} and validate it against the measurements reported in
the attestation report provided by the cloud provider.
If the measurements do not match, the \cvm{} is likely booted
without the \svsm{} firmware.

\paragraph{Weaknesses in random number generator~(RNG)}
A weak HWRNG not only poses threat for the \vtpm{} implementation, but also
for the software running inside the confidential VM.
Failing to seed the random number generator of a confidential VM correctly
can result in cryptographic key leakage\cite{kelsey1998cryptanalytic},
particularly in well documented random input signature algorithms like
ECDSA\cite{johnson2001elliptic}.
Furthermore, all \vtpm{}s require a secure RNG to operate correctly because
of their reliance on it for the generation of ephemeral keys and nonces for
secure functions.
The problem is particularly acute for an ephemeral \vtpm{} because the TPM
manufacturing stage requires the generation of unguessable seeds which can
only be achieved if they are based on an entropy source which cannot be
influenced in any way by the host.

However, AMD hardware has suffered from a buggy HWRNG in the past where
\lstinline{RDRAND} instruction gave out a constant value instead of a
random number~\cite{amd-hwrand-bug}.
An attacker could exploit a weak or buggy HWRNG implementation to guess the
initial seeds of the \vtpm{} and create the same secret keys as the
\vtpm{}.
For example, by guessing the attestation key, one could forge TPM quotes
and break the guarantees of remote attestation.
To be resilient to such hardware bugs, we can seed the RNG with additional
sources of entropy such as the hash of a key derived by the AMD-SP upon
user's request along with the \lstinline{RDSEED} instruction.

\subsection{Case Studies}

\paragraph{Full disk encryption}
Full disk encryption~(FDE) protects the confidentiality and integrity of data
at-rest.
To prevent accidental disclosure of the secret key (e.g., disk encryption
key), it is a standard practice to encrypt the secret key~(\emph{wrap}
operation) such that it can be decrypted only by the
TPM~(\emph{unwrapping}).
The wrapping key~(i.e., the key which wraps the secret) is often the
storage root key (SRK) present in the TPM.

\begin{figure}
  \centering
  \includegraphics[width=0.89\columnwidth]{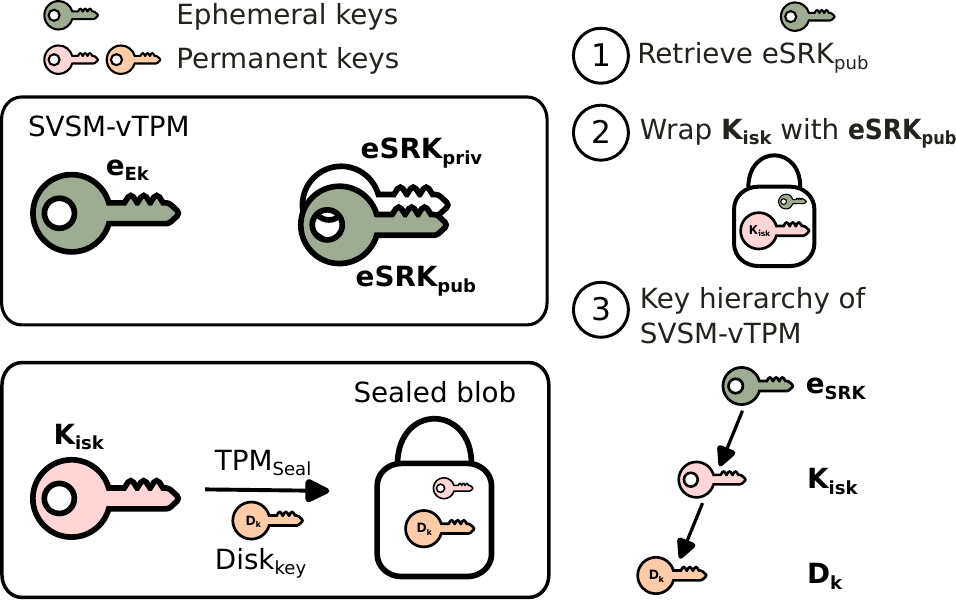}
  \caption{Full disk encryption in an ephemeral vTPM}
  \label{fig:fde}
  \vspace{-4mm}
\end{figure}

However, in our ephemeral \vtpm{}, there are no persistent storage keys in
the TPM to support unwrapping of keys.
~\autoref{fig:fde} shows the steps involved in supporting FDE on an
ephemeral \vtpm{}.
To support FDE, we create an intermediary
storage key $K_{iSK}$~(
a restricted decryption key with
sensitiveDataOrigin~\cite{tpm:sensitive-data-origin}).
Now, we perform a TPM \emph{seal} operation on the disk encryption key by
parenting it to the storage key ($K_{iSK}$) we just created, outputting a
sealed blob which can be unsealed only by a TPM with the same key.
On platform boot, the \vtpm{} would generate an ephemeral endorsement key
($e_{Ek}$) and an ephemeral storage root key~($eSRK$).
By retrieving the public part of the eSRK~(~\circlednum{1} in
~\autoref{fig:fde}), we can wrap the intermediary key
$K_{iSK}$ with $eSRK_{pub}$ to create a wrapped key that can be decrypted
only by our \vtpm{}~(~\circlednum{2} in \autoref{fig:fde}).
It has to be noted that all the above operations can be performed on any
TPM, i.e., the user need not necessarily perform these on the \vtpm{} of
the \cvm{}.
Now, the disk encryption key is wrapped to the parent key and the parent is
in turn wrapped to the eSRK, forming a hierarchy under the ephemeral storage
root key~(\circlednum{3} in \autoref{fig:fde}).
It is also possible to wrap the parent key with $EK_{pub}$ instead to
create a hierarchy under the ephemeral endorsement key.

As both the disk encryption key and its parent key are wrapped for our
specific \vtpm{}, they are no longer a secret and can be delivered to the
\cvm{} in the clear.
Since the sealed disk encryption key is invariant, we can embed this into
the initrd.
Finally, we can deliver the wrapped parent key~(\circlednum{2} in \autoref{fig:fde})
to the \cvm{} once we have performed the initial attestation of the
platform to ensure its trustworthiness.

\paragraph{Storing secrets}
We cannot store secrets directly by wrapping the keys on our ephemeral
\svtpm{} as the EK and SRK would be newly generated on every boot.
One could use a similar technique we used for FDE to form a hierarchy of
keys under an intermediary storage key.
Once the system is booted, we can parent the intermediary key to the
ephemeral SRK or EK forming a hierarchy under the chosen key.
Using this technique, one could store a hierarchy of keys, as we do with a
regular persistent TPM.

\section{Conclusions}

The landscape of cloud security is changing with the growing need to  
remove the cloud provider from the trust domain. 
Hardware vendors lay the foundation for implementing this vision through a
collection of mechanisms that ensure confidentiality of a cloud execution,
i.e.,  encryption of application memory, but, unfortunately, lack support for
ensuring runtime integrity. 
Our work develops a novel approach for virtualizing the hardware
root-of-trust through a combination of hardware mechanisms and a new
ephemeral approach to managing the TPM state.
We demonstrate how an ephemeral \vtpm{} can be used for providing
remote attestation of a \cvm{}.
In the spirit of transparency, our implementation is open
source and can be audited, verified, and extended.
As more and more cloud providers are gearing up to offer confidential
VMs, we believe our \svtpm{} architecture would provide a reference
point for implementing a \vtpm{} on \snp{} infrastructure.
While our implementation is tied to \snp{}, there is no real reason
why it could not be replicated in other TEEs. The only requirement is
to deploy ephemeral \vtpm{}s inside a secure enclave that is isolated
from the rest of the hardware (i.e. encrypted) and the guest
(the equivalent of what VMPLs provide).

\bibliographystyle{plain}

\bibliography{atc}

\end{document}